\documentclass[10pt,conference]{IEEEtran}
\IEEEoverridecommandlockouts\IEEEpubid{\makebox[\columnwidth]{ 978-1-6654-3540-6/22~\copyright~2022 IEEE \hfill} \hspace{\columnsep}\makebox[\columnwidth]{ }}
\PassOptionsToPackage{draft}{hyperref}
\usepackage{multirow}
\usepackage{cite}
\usepackage{amsmath,amssymb,amsfonts}
\usepackage{graphicx}
\usepackage{textcomp}
\usepackage{xcolor}
\usepackage{mathtools}
\usepackage{commath}
\usepackage{algorithm}
\usepackage{algpseudocode}
\usepackage{hyperref}
\hypersetup{colorlinks,allcolors=black}
\usepackage[compact]{titlesec}
\titlespacing{\section}{0pt}{2ex}{1ex}
\titlespacing{\subsection}{0pt}{1ex}{0ex}
\titlespacing{\subsubsection}{0pt}{0.5ex}{0ex}
\usepackage{amsmath}

\def\BibTeX{{\rm B\kern-.05em{\sc i\kern-.025em b}\kern-.08em
    T\kern-.1667em\lower.7ex\hbox{E}\kern-.125emX}}
    
\begin{document}
 
\title{Physics-Inspired Deep Learning Anti-Aliasing Framework in Efficient Channel State Feedback}
\author{Yu-Chien Lin, Yan Xin, Ta-Sung Lee, Charlie (Jianzhong) Zhang, and Zhi Ding
\thanks{Y.-C Lin and Z. Ding are with the Department of Electrical and Computer Engineering,
University of California, Davis, CA, USA (e-mail: ycmlin, zding@ucdavis.edu).

Y. Xin and C. Zhang are with Samsung Research America, USA (e-mail: yan.xin, jianzhong.z@samsung.com).

T.-S Lee is with the Institute of Communications Engineering, National Yang Ming Chiao Tung University, Taiwan (e-mail: tslee@nycu.edu.tw).}

\thanks{This work is based on materials supported by the National Science Foundation under Grants 2029027 and 2002937 (Lin, Ding) and by the National Science and Technology Council of Taiwan under grant NSTC 112-2218-E-A49-020, NSTC 112-2221-E-A49-079 (Lee and Lin).}}

\maketitle
\pagestyle{empty}
\thispagestyle{empty}
\begin{abstract}
Acquiring downlink channel state information (CSI) at the base station is vital for optimizing performance in massive Multiple input multiple output (MIMO) Frequency-Division Duplexing (FDD) systems. While deep learning architectures have been successful in facilitating UE-side CSI feedback and gNB-side recovery, the undersampling issue prior to CSI feedback is often overlooked. This issue, which arises from low density pilot placement in current standards, results in significant aliasing effects in outdoor channels and consequently limits CSI recovery performance. To this end, this work introduces a new CSI upsampling framework at the gNB as a post-processing solution to address the gaps caused by undersampling. Leveraging the physical principles of discrete Fourier transform shifting theorem and multipath reciprocity, our framework effectively uses uplink CSI to mitigate aliasing effects. We further develop a learning-based method that integrates the proposed algorithm with the Iterative Shrinkage-Thresholding Algorithm Net (ISTA-Net) architecture, enhancing our approach for non-uniform sampling recovery.
Our numerical results show that both our rule-based and deep learning methods significantly outperform traditional interpolation techniques and current state-of-the-art approaches in terms of performance.

\end{abstract}

\section{Introduction}
Massive multiple-input multiple-output (MIMO) improves spectrum and energy efficiency in wireless systems, but requires accurate downlink (DL) channel state information (CSI) acquisition at the base station or gNodeB (gNB). In frequency-division duplexing (FDD) systems, DL CSI acquisition depends on UE feedback, which can be costly due to its large number of channel coefficients. Efficient compressive CSI feedback is crucial to conserve uplink (UL) bandwidth and UE power for practical deployment of massive MIMO in FDD wireless networks.

Cellular CSI has a limited delay spread, which is a characteristic of radio physics. Efficient user equipment (UE) feedback can take advantage of this delay spread sparsity to compress CSI. One approach to efficient CSI compression and recovery is the use of a deep autoencoder framework, as demonstrated in \cite{CsiNet}. This framework includes an encoder at the UE and a decoder at the serving gNB. Other related works have also demonstrated superior CSI recovery or lightweight design using various autoencoder models, such as \cite{CsiNet-LSTM, CRNet, DeepCMC, CLNet}. In addition to autoencoders, more recent works have utilized the underlying channel correlation to aid and improve the recovery of DL CSI at BSs. These approaches include using previous CSI \cite{CsiNet-LSTM, MarkovNet}, CSI of nearby UEs \cite{CoCsiNet}, and UL CSI \cite{CQNET, DualNet, DualNet-MP}. Further advances have focused on reducing model complexity and storage size to facilitate practical and low-cost deployment of DL-based CSI compressive feedback architecture in wireless networks, as demonstrated in \cite{CRNet, CLNet}.

A notable issue often overlooked in the previous CSI feedback framework is the implicit assumption of dense pilot placement. It is presumed that high-quality CSI feedback leads to accurate recovery of the full CSI. Yet, the primary function of CSI reference signals (CSI-RS) in the current 5G standard, as referenced in \cite{3GPPtypeI, 3GPPtypeII}, is tailored for subband-level precoder index feedback, overlooking detailed CSI structural information. The recovery of full DL CSI, specifically at the subcarrier-level DL CSI, necessitates significantly higher pilot placement density. This increased density is crucial to effectively capture the rapid frequency domain variations due to large-delay multipaths. This factor is fundamental in explaining why normalized mean square error (NMSE) performance typically degrades more in outdoor channels, which have a higher presence of large-delay multipaths, compared to indoor scenarios. However, adhering to the current standard's CSI-RS placement density introduces severe aliasing effects, hindering the recovery of the full DL CSI. To our knowledge, there has been no previous research dedicated to addressing this aliasing problem in CSI feedback.\\

Super-resolution (SR), as detailed in \cite{super-resolution-background}, is a captivating technique in computer vision that focuses on enhancing image resolution. This technique involves upsampling low-resolution (LR) images to provide more detailed and clearer elements, such as edges, thereby improving the overall image quality. In \cite{SRCNN}, a SR-CNN has been shown to achieve superior performance by applying a convolutional neural networks to interpolation.   More recently, Gao et al. in \cite{Gao_2023_CVPR} implicit diffusion model (IDM), a method for upsampling LR images using implicit neural representation alongside a denoising diffusion model. Additionally, Fang et al., as cited in \cite{Fang_2022_CVPR}, developed a hybrid network combining CNN and transformer technology, offering a comparatively lightweight solution relative to IDM. However, it is important to note that these methods are not directly applicable to CSI upsampling, due to the inherent differences between the nature of CSIs and images.\\

From the perspective of signal processing and information theory, the loss of high-frequency variations during sampling often makes it impossible to reconstruct the original signals from its downsampled version, due to the information gap that arises. In computer vision, deep learning models effectively learn extensive prior information from images to bridge this gap, leveraging common features such as facial features, colors, textures, edges, and shapes. For instance, when a deep learning model identifies a specific patch as a face, it greatly reduces the uncertainty in upsampling LR images, as it expects only facial features in that area. However, this approach differs from the SR task in computer vision, as the details in CSIs are random and more challenging to be learned as prior information for a deep learning model. To address this, we propose utilizing UL CSI information to counteract aliasing effects due to insufficient pilot sampling rate, by exploiting multipath reciprocity. This approach aims to fill the information gap inherent in the CSI upsampling process.

Our primary objective is to address the undersampling issue caused by CSI-RS pilot placement in CSI feedback of the existing cellular network standard. We introduce a CSI upsampling methodology utilizing UL CSI, which assists in designing a bandpass filter to mitigate the undersampling problem. We meticulously craft a physics-inspired deep learning architecture that leverages UL CSI for effective aliasing suppression. Our key contributions can be summarized as follows:

\begin{itemize}
    \item We develop a low-complexity and rule-based technique, termed UL Masking, which leverages DFT shifting theorem in uniformly sampled signals and multipath reciprocity to create a bandpass filter that suppresses aliasing peaks.
    
    \item We establish a deep learning framework, SRCsiNet, that unfolds and expands the UL Masking approach. This framework enhances the utilization of the DFT shifting theorem and multipath reciprocity.
    
    \item We train our framework end-to-end, compelling the non-aliasing selection map generation module to construct an effective bandpass filter for aliasing suppression. This filter is then implemented in the subsequent CSI attention and refinement module within the beam-delay domain.
    
    \item We introduce a novel CSI upsampling strategy that integrates the strengths of ISTA-Net and our proposed SRCsiNet. This approach facilitates non-uniform sampling recovery and proficient aliasing suppression.
\end{itemize}

%%%%%%%%%%%%%%%%%%%%%%%%%%%%%%%%%%%%%%%%%%%%%%%%%%%%%%%%%%%%%%%%%%%%%%
\section{System Model}
\subsection{DL CSI Preprocessing}

We consider a single-cell MIMO FDD link where a gNB with $N_a$ antennas 
serves a plurality of single-antenna UEs. 
Following 3GPP technical specifications, sparse pilot symbols (i.e., CSI-RS) are uniformly distributed in frequency domain for DL channel acquisition. Assuming each subband contains $N_f$ subcarriers with a spacing of ${\Delta}f$ and a pilot spacing of $D_\text{RS}$ subcarriers, 
adjacent CSI-RSs are seperated by $D_\text{RS}{\cdot}{\Delta}f$ Hz. We denote $\mathbf{h}_{i} \in \mathbb{C}^{M_f \times 1}$ as CSI-RS DL 
CSI of the $i$-th antenna at gNB at $M_f$  
pilot positions.
Let superscript $(\cdot)^H$ denote the conjugate transpose. By collecting 
CSI of each gNB, a pilot sampled DL CSI matrix $\mathbf{H}_\text{RS}$ relates to the full DL CSI matrix ${\mathbf{H}} \in \mathbb{C}^{N_a \times N_f}$ via
\begin{equation*}
    \mathbf{H}_\text{RS} = {\mathbf{H}}\mathbf{Q}_{D_\text{RS}} =
    \begin{bmatrix}
    \mathbf{h}_1\;
    \mathbf{h}_2\;
    \cdots\;
    \mathbf{h}_{N_a}
    \end{bmatrix}^H \in \mathbb{C}^{N_a \times M_f},
    \label{DL pilot CSI matrix}
\end{equation*}
where $\mathbf{Q}_{D_\text{RS}} = [\mathbf{e}_1,\mathbf{e}_{1+D_\text{RS}},...,\mathbf{e}_{1+(M_f-1)D_\text{RS}}]\in\mathbb{C}^{N_f\times M_f}$ is a downsampling matrix with pilot rate $D_\text{RS}$ with $\mathbf{e}_i \in \mathbb{C}^{N_f}$ being the $i$-th column vector of an identity matrix of size $N_f$. 

%To reduce feedback overhead, we
%exploit physical angular and multipath delay sparsity of CSI by transforming full DL CSI into angle-delay (AD) domain through discrete Fourier transform (DFT) or discrete cosine transform (DCT). We then truncate the insignificant near-zero elements in trailing delay indices as follows:
%\begin{equation}
%    \mathbf{M} = \widetilde{\mathbf{H}} \cdot \mathbf{F}_D\cdot 
%    \underbrace{\left[\begin{array}{c} \mathbf{I}_{N_t\times N_t}\\
%\mathbf{0}\end{array}\right]}_{\mathbf{T}} \in \mathbb{C}^{N_a \times N_t},
%    \label{CSI_domain}
%\end{equation}
%where $\mathbf{F}_D \in \mathbb{C}^{M_f \times M_f}$ denotes a sparse transformation matrix such as IDFT or DCT matrix whereas
%$\mathbf{T}$ denotes delay domain truncation. Note that the matrix $\mathbf{T}$ may be
%controlled according to transformation and CSI properties. Matrix $\mathbf{T}$ in 
%Eq.~(\ref{CSI_domain}) is an example for DCT transformation that drops the last $M_f-N_t$ columns of $\widetilde{\mathbf{H}} \cdot \mathbf{F}_D$ corresponding to long (but negligible)
%multipath delays.
% \footnote{Usually, CSIs are normalized before compression and recovery. In this work, we conduct a global normalization to make element value ranging from -1 to 1.}

\subsection{DL CSI Feedback}
Autoencoder has shown successes for CSI compression. An encoder at UE compresses its estimated DL CSI based on reference signals for UL feedback and a decoder at gNB recovers the CSI according to the feedback from UE. Before compression and after recovery, some works \cite{CsiNet, SCEnet} may or may not transform CSI into the domain with sparse features as pre-processing, which usually only pose slight impact. Many have exploited convolutional and fully connected layers to compress and recover the DL pilot CSI via
\begin{align*}
\mbox{Encoder:}\quad& \mathbf{q} = f_\text{en}(\mathbf{H}_\text{RS} + \mathbf{N}), \\
\mbox{Decoder:}\quad&
    \widehat{\mathbf{H}}_\text{RS} = f_\text{de}(\mathbf{q}). 
\end{align*}
We note that the size of the codeword $\mathbf{q} \in \mathbb{C}^{\frac{N_aM_f}{CR}}$ for UL feedback is determined by a specific compression ratio $CR$. We can evaluate the feedback loss by the NMSE of the pilot DL CSI:

\begin{equation*}
    \textit{Loss}_\text{FB}(\widehat{\mathbf{H}}_\text{RS}, \mathbf{H}_\text{RS}) = \sum_{d=1}^{D} \frac{\norm{\widehat{\mathbf{H}}_{\text{RS},d} -\mathbf{H}_{\text{RS},d}}^2_{F}}{\norm{\mathbf{H}_{\text{RS},d}}^2_{F}}, \label{Eq: loss feedback}
\end{equation*}
where subscript $d$ denotes the $d$-th random test.

\subsection{Aliasing Issue}
The gNB now designs its precoder based on the full-size DL CSI, moving away from the limited CSI-RS DL CSIs. Our primary interest shifts towards the total discrepancy between the actual full DL CSI, denoted as $\mathbf{H}$, and the estimated full DL CSI, denoted as $\widehat{\mathbf{H}}$. The discrepancy is given as follows:

\begin{equation*}
    \textit{Loss} = \text{NMSE}(\widehat{\mathbf{H}}, \mathbf{H}) = \sum_{d=1}^{D} \frac{\norm{\widehat{\mathbf{H}}_d-{\mathbf{H}}_d}^2_{F}}{\norm{{\mathbf{H}}_d}^2_{F}}, \label{Eq: loss upsampling}
\end{equation*}
\begin{equation*}
    \widehat{\mathbf{H}} = f_\uparrow(f_\text{de}(f_\text{en}(\mathbf{H}_\text{RS} + \mathbf{N}))), \label{Eq: H_hat}
\end{equation*}
where $f_\uparrow(\cdot)$ is the upsampling operation and $\widehat{\mathbf{H}} \in \mathbb{C}^{N_a \times N_f}$ is the estimated DL CSI after upsampling/interpolation.

\begin{figure}
    \centering
    \resizebox{3.4in}{!}{
    \includegraphics{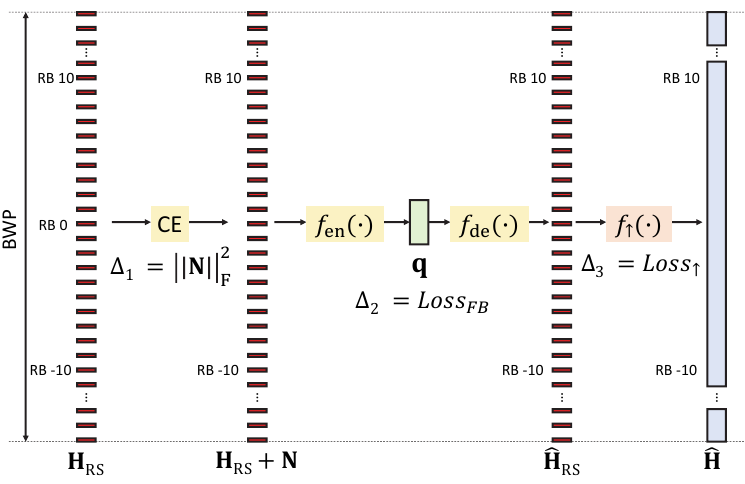}}
    \caption{Illustration of the total discrepancy related to the losses at different stages. ${\Delta}_1$, ${\Delta}_2$ and ${\Delta}_3$ denote the distortions from channel estimation at UE side, feedback from UE to gNB, and upsampling, respectively.}
    \label{fig: dlcsiloss}
\end{figure}

As shown in Fig. \ref{fig: dlcsiloss}, the total discrepancy in recovering the full DL CSI, denoted as $Loss$, arises from three main factors: channel estimation (CE) noise $\mathbf{N}$, feedback loss $Loss_\text{FB}$, and upsampling/interpolation loss $Loss_\uparrow$. The CE loss, resulting from imperfect CE at the UE side, has been effectively addressed by rule-based methods like Least Square (LS) and MMSE estimation \cite{denoising1}, as well as advanced learning-based denoising networks \cite{denoising2, denoising3}. Feedback loss, due to limited CSI feedback, has been extensively explored in existing CSI feedback frameworks \cite{CsiNet, DualNet}. However, there has been less focus on upsampling loss. This loss occurs when interpolating full DL CSIs from a limited number of known estimated pilot DL CSIs. While feedback loss $Loss_\text{FB}$ is typically predominant in indoor propagation channels, the insufficient density of current CSI-RS placements means that upsampling loss $Loss_\uparrow$ becomes a significant challenge in recovering DL CSIs with large delay spread (i.e., fast-varying in frequency domain).  
%%%%%%%%%%%%%%%%%%%%%%%%%%%%%%%%%%%%%%%%%%%%%%%%%%%%%%%%%%%%%%%%% 
% to modify the equation and its corresponding description

Prior research often assumes adequate pilot density in the frequency domain for all types of channels. However, the density of pilot placement in CSI-RS, as specified in cellular network standards \cite{3GPP}, falls short for outdoor scenarios, particularly for channels with a high delay spread. This leads to a significant issue: the CSI-RS DL CSI matrix, $\mathbf{H}_\text{RS}$, may experience aliasing due to downsampling, rendering it impossible to accurately recover the full DL CSI, $\mathbf{H}$.
Let us define the pilot sample rate in frequency as $S_\text{F}$ and the maximum delay tap as ${\Delta}t_\text{max}$ seconds. If $\frac{1}{2S_\text{F}} \leq {\Delta}t_\text{max}$, the channels captured from CSI-RS are considered to be aliased signals. Generally, recovering aliased signals (i.e., aliased downsampled (DS) CSI) to their original form (i.e., full CSI) is not feasible. However, if the DS signals satisfy certain constraints, we may recover the full CSI with aids of side information, which will be introduced in the following sections. \\

Previous studies often assume an overly idealistic approach to upsampling/interpolation, which can be a \textit{critical operation} in channels with a large delay spread, and results in a bottleneck in reducing the total discrepancy\footnote{As the three operations (estimation, feedback, and interpolation) are sequential, the one causing the largest loss becomes the bottleneck in reducing the total discrepancy. This operation is termed the critical operation.}. To enhance the overall performance, our focus should shift to improving this critical operation rather than the other two.\\

\section{UL-CSI aided Upsamplign with Aliasing Suppression}

\subsection{CSI Upsampling with Side Information}
For an arbitrary channel $\mathbf{H} \in \mathbb{C}^{N_a \times N_f}$ in frequency domain and its DS version $\mathbf{H}_\text{RS} = \mathbf{H}\mathbf{Q}_{D_\text{RS}} \in \mathbb{C}^{N_a \times M_f}$ by a factor of $D_\text{RS}$. If we upsample the $\mathbf{H}_\text{RS} $ by inserting $D_\text{RS}-1$ zeros between any two consecutive samples along frequency domain, we have
\begin{equation}
    \mathbf{H}_\text{DS}[:,j] = \left\{ 
    \begin{aligned}
    &\mathbf{H}[:,j], &\forall j \in \Psi_\text{RS},\\
    &\mathbf{0}, &\forall j \notin \Psi_\text{RS},   
    \end{aligned}
    \right.
    \label{eq: repmat1}
\end{equation}
where $\Psi_\text{RS} = \{0,D_\text{RS},...,(M_f-1)D_\text{RS}\}$ is a downsampling index set. Note that $\mathbf{H}_\text{DS}$ consists of the entries of $\mathbf{H}_\text{RS}$ at frequencies with pilots and zeros elsewhere. By DFT/IDFT transformation, the full and DS DL CSI in beam-delay (BD) domain can be obtained as follows:
\begin{equation*}
    \mathbf{H}_\text{BD} = \mathbf{F}_\text{AB}\mathbf{H}\mathbf{F}_\text{FD} \in \mathbb{C}^{N_a \times N_f},
\end{equation*}
    
\begin{equation}
    \mathbf{H}_\text{DS,BD} = \mathbf{F}_\text{AB}\mathbf{H}_\text{DS}\mathbf{F}_\text{FD} \in \mathbb{C}^{N_a \times N_f},
    \label{eq: repmat2}
\end{equation}
where $\mathbf{F}_\text{AB} \in \mathbb{C}^{N_a \times N_a}$ and $\mathbf{F}_\text{FD} \in \mathbb{C}^{N_f \times N_f}$ are DFT and IDFT transformation matrices, respectively. The subscripts AB and FD denote the transformation from antenna/frequency to beam/delay domains, respectively. Note that we use subscript BD, AD, AF to denote CSI in beam-delay, angle-delay, and angle-frequency domains, respectively. We use no subscript to denote CSI in the orignal domain which is antenna-frequency domain.

\begin{figure}
    \centering
    \resizebox{3.4in}{!}{
    \includegraphics{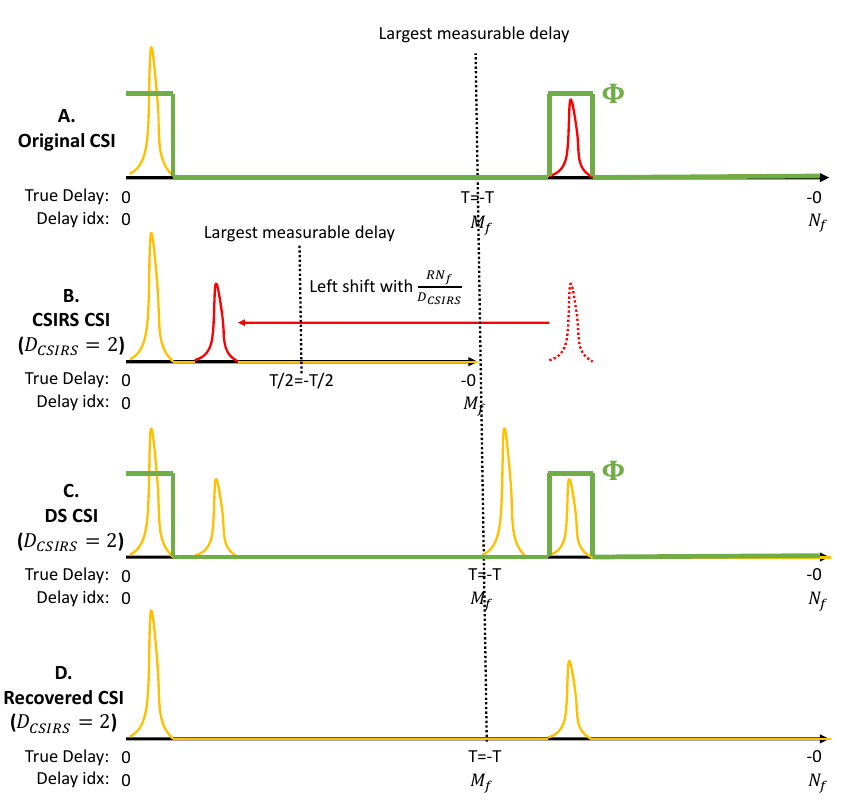}}
    \caption{Illustration of CSI upsampling with side information. (A) shows the original CSI magnitude in delay domain. (B) demonstrates the CSIRS CSI magnitude in delay domain when $D_\text{RS} = 2$. We can find that the high negative delay peak wraps around ($R=1$) into the low delay region, leading aliasing effect. (C) shows the DS CSI magnitude in delay domain by inserting zero inbetween samples of CSIRS CSI in frequency domain. The green curve represents an ideal binary bandpass filter $\mathbf{\Phi}$ to be the side information. (D) is the resulting DL CSI magnitude in delay domain after applying the binary bandpass filter $\mathbf{\Phi}$.}
    \label{fig: biniso}
\end{figure}

Given the \textit{DFT shifting theorem}\cite{DFT_shifting_theorem}, after IDFT transformation, we have the following relationship between the full and DS DL CSIs:

\begin{equation}
    \begin{aligned}
        &\mathbf{H}_\text{DS,BD}[i,j] = \\
        &\left\{
        \begin{aligned}
            &\frac{\begin{aligned}
            \mathbf{H}_\text{BD}[i,j]+\mathbf{H}_\text{BD}[i,j+M_f]+\\
            ...+\mathbf{H}_\text{BD}[i,j+M_f(D_\text{RS}-1)]
            \end{aligned}}{D_\text{RS}}, \forall 0 \leq j < M_f\\
            &\mathbf{H}_\text{DS,BD}[i,\text{mod}(j,M_f)], \text{otherwise}  
        \end{aligned}\right..
    \end{aligned}
    \label{eq:aliasing}
\end{equation}
Note that $\mathbf{H}_\text{DS,BD}$ is periodic in the delay domain with a period of $M_f = N_f/D_\text{RS}$. If $\mathbf{H}_\text{BD}[i,j] \neq 0$ for any $j > M_f$, we can say that the aliasing effect occurs and it cannot be recovered to the original version $\mathbf{H}$ in general cases since we can only measure $\mathbf{H}_\text{DS,BD}$, the sum of the multipaths. However, since $\mathbf{H}_\text{DS,BD}$ is periodic in the delay domain with a period of $M_f$, which matches the wrapped-around effect due to downsampling, the IDFT transformation \textit{unwraps} the delay bins of $\mathbf{H}_\text{BD}$ to the original delay positions. Thus, $\mathbf{H}$ can be recovered if $\mathbf{H}_\text{BD}[i,j]$ in the delay domain satisfies the two requirements shown below:
\begin{itemize}
    \item \textbf{Bin Isolation Property}: for any non-zero $\mathbf{H}_\text{DS,BD}[i,j]$ in Eq.(\ref{eq:aliasing}), only one from the $D_\text{RS}$ aliased copies $\mathbf{H}_\text{BD}[i,j], \mathbf{H}_\text{BD}[i,j+N_f/D_\text{RS}],..., \mathbf{H}_\text{BD}[i,j+N_f(D_\text{RS}-1)/D_\text{RS}]$ is non-zero. Namely, the delay bins (i.e., $\mathbf{H}_\text{BD}[i,j], j > M_f$) and the low-delay bin (i.e., $\mathbf{H}_\text{BD}[i,j], j \leq M_f$) are isolated after wrapped-around in its DS version. If the bin isolation property holds, each non-zero DS signal $\mathbf{H}_\text{DS,BD}[i,j]$ in delay domain maps to a scaled unique delay bin in the original signal (i.e., $\mathbf{H}_\text{DS,BD}[i,j] = \mathbf{H}_\text{BD}[i,n_k]/D_\text{RS}$). Note that $n_k$ can only be $j$, $j + M_f$,..., or $j + (D_\text{RS}-1)M_f$.
    \item \textbf{Knowledge of bin locations}: we have the perfect knowledge map $\mathbf{\Phi} \in \mathbb{C}^{N_a \times N_f}$ with ones at the positions with non-zero values in the the full CSI matrix $\mathbf{H}_\text{BD}[i,j]$ and zeros elsewhere.
\end{itemize}
Fig. \ref{fig: biniso} shows a simple illustration for the single antenna case with the intermediate results of the proposed CSI upsampling approach using the bin location information. If the full CSI matrix $\mathbf{H}_\text{BD}$ satisfies the above two requirements, $\mathbf{H}_\text{BD}$ can be ideally obtained by
\begin{equation*}
    \widehat{\mathbf{H}}_\text{BD} = D_\text{RS}\mathbf{\Phi}\circ\mathbf{H}_\text{DS,BD} \approx \mathbf{H}_\text{BD} .
\end{equation*}

Note that $\circ$ denotes the element-wise product operation. $\mathbf{\Phi}$ acts like a bandpass filter in BD domain. Although the two requirements are ideal, they lead us to a rationale to deal with aliasing problems. That is, to deal with sparse signals, we can suppress aliasing peaks with the knowledge of the non-zero bin locations as a \textit{bandpass filter}. In practice, DL CSI is somehow sparse so that a \textit{quasi bin isolation property} can hold. As for the knowledge of bin locations of DL CSI, we can estimate it according to UL CSI at base stations.

\subsection{Multipath Reciprocity}
\begin{figure}
    \centering
    \resizebox{3.4in}{!}{
    \includegraphics{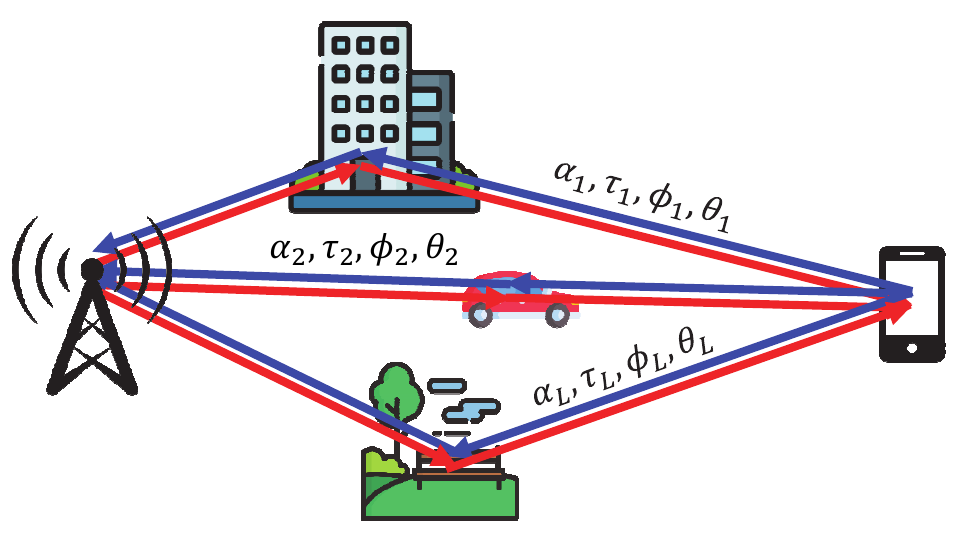}}
    \caption{Illustration of multipath reciprocity between UL and DL propagation channels.}
    \label{fig:FDD Reciprocity}
\end{figure}

Typically, acquiring the exact delay bin location information without the original DL CSI, denoted as $\mathbf{H}_\text{BD}$, is challenging. However, in communications systems, the DL CSI $\mathbf{H}_\text{BD}$ is often closely correlated with the UL CSI, which is readily available at base stations, especially in terms of magnitudes in the BD domain. Although DL and UL CSIs do not exhibit full correlation in FDD wireless systems, as illustrated in Fig. \ref{fig:FDD Reciprocity}, they often share similar large-scale multipath geometries. This multipath reciprocity results in comparable delay and angle profiles, a finding supported by field tests and mathematical analysis \cite{prof_mr, prof_mr1}. Therefore, UL CSI in the BD domain is typically considered a reliable estimate for the AD profiles of DL CSI. Owing to the relatively high pilot placement density in UL CSI, there are no aliasing effects, allowing for the design of a bandpass filter to mitigate aliasing effects in DL CSIs.

\begin{figure}
    \centering
    \resizebox{2.5in}{!}{
    \includegraphics{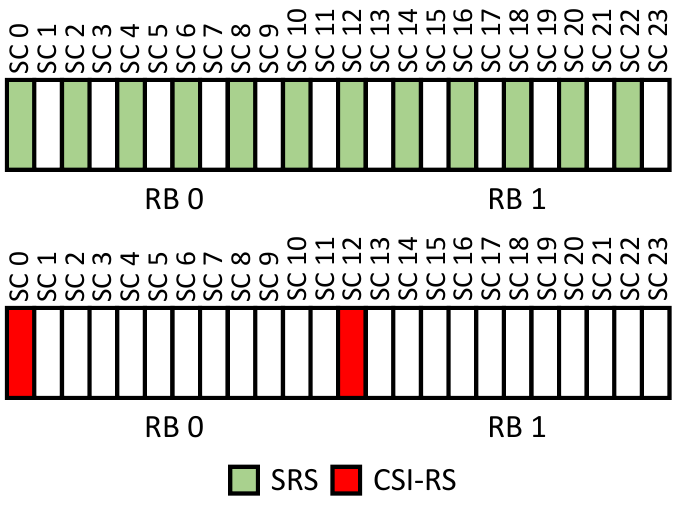}}
    \caption{Comparison of SRS and CSI-RS placement density.}
    \label{fig: illus_SRS_CSIRS}
\end{figure}

In modern communication systems, as depicted in Fig. \ref{fig: illus_SRS_CSIRS}, the pilot placement density in the frequency domain of the Sounding Reference Signal (SRS) is much higher (every two subcarriers) compared to that of CSI-RS (every 12 subcarriers). Consequently, the maximum non-aliasing delay (i.e., measurable delay) of UL CSI is approximately six times greater than that of DL CSI, virtually eliminating aliasing effects in UL CSIs. Based on the principle of multipath reciprocity, this work proposes designing the bandpass filter  $\mathbf{\Phi}$ using UL CSI information.

\subsection{UL Masking: UL-Assisted CSI Upsampling with Aliasing Suppression}
Assume that we have perfect UL CSI $\mathbf{H}_\text{UL}$. According to the multipath reciprocity between UL and DL CSIs, we can design a two-dimensional bandpass filter based on the UL CSI magnitude in BD domains as follows:

\begin{equation*}
    \mathbf{\Phi}_\text{UL}[i,j] = \left\{
    \begin{aligned}
        &0, & |\mathbf{H}_\text{UL,BD}[i,j]| <  T,\\
        &1, & |\mathbf{H}_\text{UL,BD}[i,j]| \geq  T,
    \end{aligned}
    \right.
\end{equation*}

\begin{equation*}
    \mathbf{H}_\text{UL,BD} = \mathbf{F}_\text{AB}\mathbf{H}_\text{UL}\mathbf{F}_\text{FD} \in \mathbb{C}^{N_a \times N_f},
\end{equation*}
where we set $T = R\cdot \sqrt{P}$ and $P$ is the average power of $\mathbf{H}_\text{UL,BD}$.
We next can estimate the BD domain DL CSI by
\begin{equation*}
    \widehat{\mathbf{H}}_\text{BD} = \mathbf{\Phi}_\text{UL}\circ\mathbf{H}_\text{DS,BD}.
\end{equation*}

Due to the multipath reciprocity, the filter can effectively suppress the aliased copies as long as we design a proper threshold $T$ which determines the pass band in delay and angle domains. However, it is challenging to find a reasonable threshold $T$ for all CSIs. 
 
\section{Physic-inspired AI-driven Aliasing Suppression}
Previous works \cite{CANet,CLNet,CoCsiNet} have been successfully applied to  in CSI compression and recovery. Enough pilot sampling rate was usually assumed. In fact, following the 3GPP 5G NR standard \cite{3GPP}, UEs estimate the channels from CSI-RS and send channel state feedback. However, the frequency density of CSI-RS is not sufficient to capture the fast channel variation along frequency domain. Even if a perfect CSI feedback is achieved, the aliasing loss due to downsampling is theoretically not possible to be recovered.

\subsection{\textbf{Model Architecture}}
There are plenty of successful network architecture which  can enhance image details while maintaining visual fidelity after SR operation. In a sense of information theory, the model learns prior information from the training data to fill the information gap between the target and desired images. There are lots of common features in images such as facial features, colors textures, edges and shapes. For example, as long as the deep learning model can recognize a specific patch as a face, it can largely lower the uncertainty to upsample the LR images since there exists nothing else except facial features. However, unlike SR task in computer vision, the details of CSIs are random and difficult to learn as prior information stored in the deep learning model. To fill the information gap, we propose to utilzie UL CSI information by exploiting multipath reciprocity against aliasing effects due to an insufficient pilot sampling rate.

This section introduces a general learning framework designed to effectively upsample LR tensors into SR equivalents. This process is akin to the SR challenge in computer vision, where numerous successful networks \cite{SRCNN, Gao_2023_CVPR, Fang_2022_CVPR} have been developed to enhance image details while preserving visual fidelity after SR operation. From the perspective of information theory , the model employs prior knowledge obtained from training data to fill the gap between actual and desired images. Certain image features, including facial characteristics, colors, textures, edges, and shapes, are common across various images. These features are retained as prior knowledge within the model, ready to be utilized as necessary to aid in image processing tasks. For instance, if a deep learning model identifies a particular segment as part of a face, it significantly reduces the uncertainty involved in upscaling LR images, since the expected features are confined to those associated with faces.\\

\begin{figure}
    \centering
    \resizebox{2.5in}{!}{
    \includegraphics{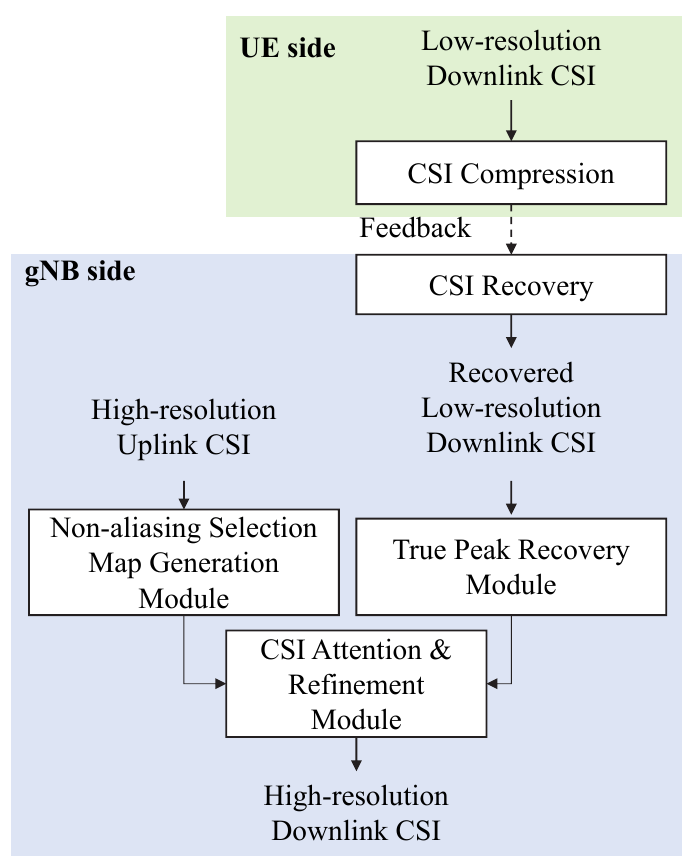}}
    \caption{General architecture of the proposed physic-inspired AI-driven aliasing suppression framework. This framework consists of two parts. The first part is CSI compression and recovery which are deployed at UE and base station sides, respectively. The other part is the SR operation for the LR CSIs.}
    \label{fig:GeneArch}
\end{figure}

However, unlike the SR task in computer vision, the intricacies of CSI are random and challenging to learn as pre-existing information within a deep learning model. To overcome this information gap, we propose leveraging UL CSI data, exploiting the principle of multipath reciprocity to counteract the aliasing effects stemming from an inadequate pilot sampling rate. Fig. \ref{fig:GeneArch} gives a high-level understanding of the proposed architecture. This framework is designed to be deployed at base stations and consists of three modules: a) non-aliasing selection map generation, b) true peak recovery, and c) CSI attention and refinement which are described in detail as follows:

\subsubsection{\textbf{True Peak Recovery}}
This module aims to upsample LR DL CSIs by inserting zeros and transform them into the beam and delay domains. By doing so, we can have a DL CSI map in BD domain which is periodic in delay domain. According to the DFT shifting invariance property, we can map the aliasing delay bins to its original positions by inserting $D-1$ zeros in between samples. On the other hand, this will also lead to more false peaks in the repetition map at the false delay positions. To implement, we basically follow Eqs. (\ref{eq: repmat1}) and (\ref{eq: repmat2}) to generate the desired repetition map $\mathbf{H}_\text{BD,DS}$. We describe these operations as a linear function $f_\text{TPR}(\cdot)$ such that $\mathbf{H}_\text{BD,DS} = f_\text{TPR}(\mathbf{H}_\text{RS})$.

\begin{figure}
    \centering
    \resizebox{3.4in}{!}{
    \includegraphics{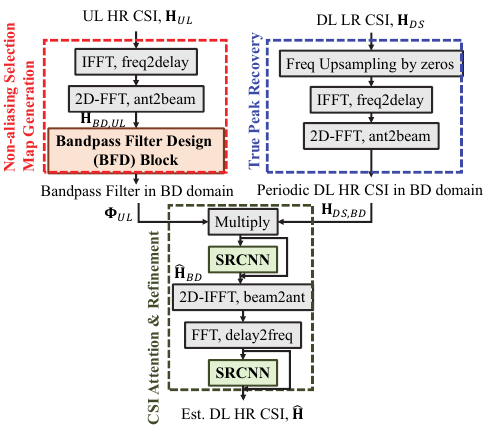}}
    \caption{Network architecture of SRCsiNet. It consists of three modules: 1) Non-aliasing selection map generation, 2) True peak generation and 3) CSI attention and refinement.}
    \label{fig:SRCsiNet}
\end{figure}
\subsubsection{\textbf{Non-aliasing Selection Map Generation (Bandpass Filter Design)}}
This module aims to generate a bandpass filter in the BD domain which can suppress aliasing peaks at wrong delay positions. Regarding the multipath reciprocity, we can reply on UL CSI to infer where the true peaks are. Instead of using a rule-based approach mentioned in the previous section, we adopt a neural network to design a bandpass filter. We first transform the HR UL CSI into BD domain as $\mathbf{H}_\text{BD,UL}$ with the same size of the matrix $\mathbf{H}_\text{BD,DS}$ to be filtered. We then feed $\mathbf{H}_\text{BD,UL}$ into three convolutional layers with two ReLU activations at the outputs of the first two convolutional layers. We then utilize a sigmoid function as the last activation function to output the bandpass filter $\mathbf{\Phi}_\text{UL}$ since it perfectly matches the soft filtering purpose (i.e., model cannot only yield zeros to suppress aliasing delay positions and ones elsewhere, but also yield values between 0 and 1 to represent the model uncertainty and provide flexibility). We called it as Bandpass Filter Design (BFD) Block. For brevity, we can express the output of the branch of the model as
\begin{equation}
    \mathbf{\Phi}_\text{UL} = f_\text{BFD}(\mathbf{H}_\text{UL}).
    \label{eq: BFD}
\end{equation}

\subsubsection{\textbf{CSI Attention and Refinement}}
This module aims to filter out the aliasing peaks and do refinement to generate the final DL CSI estimates which can be expressed as $\widehat{\mathbf{H}} = f_\text{AR}({\mathbf{\Phi}_\text{UL}\circ \mathbf{H}_\text{BD,DS}})$. The function $ f_\text{AR}(\cdot)$ aims to further refine and smooth the filtered result which may have some artifacts due to the imperfect bandpass filter $\mathbf{\Phi}_\text{UL}$ and the overlapped delay bins in $\mathbf{H}_\text{BD,DS}$. We apply two residual blocks with SRCNN block \cite{SRCNN} as the backbone to refine the estimate first in BD domain and then in AF domain.

\subsection{\textbf{Loss Function Design}}
This network aims to minimize the upsampling loss $Loss_\uparrow$ which is defined as 

\begin{equation*}
\begin{aligned}
    & Loss_\uparrow(\Theta_\text{BFD}, \Theta_\text{AR}) = \frac{1}{D}\sum_d^D \norm{\widehat{\mathbf{H}}_d - \mathbf{H}_d}_F^2, \\
    & = \frac{1}{D}\sum_d^D \norm{f_\text{AR}(\mathbf{\Phi}_{\text{UL},d}\circ \mathbf{H}_\text{BD,DS}) - \mathbf{H}_d}_F^2, \\
    & = \frac{1}{D}\sum_d^D \norm{f_\text{AR}(f_\text{BFD}(\mathbf{H}_{\text{UL},d})\circ \mathbf{H}_\text{BD,DS}) - \mathbf{H}_d}_F^2,
\end{aligned}
\end{equation*}
where $\Theta_\text{BFD}$ and $\Theta_\text{AR}$ are trainable parameters of the functions $f_\text{BFD}(\cdot)$ and $f_\text{AR}(\cdot)$, respectively.

\section{Efficient Channel State Feedback with Aliasing Suppression from Non-uniform Sampling}
The true delay position information can significantly improve the CSI recovery for high-delay scenarios. In the perspective of the information theory, if we can increase the mutual information between the input and the desired output, we can further improve the CSI recovery accuracy.

According to the 3GPP 5G-NR standards \cite{3GPP}, the primary and secondary synchronization signals (PSS and SSS) play crucial roles in cell identification and frame synchronization, appearing periodically every 25 subframes (approximately 25ms) and spanning 64-128 subcarriers in bandwidth. Beyond these primary functions, as depicted in Fig. \ref{fig: virtual_pilots}, UEs can also utilize PSS and SSS to estimate DL CSI, treating these signals as \textit{virtual pilots} for DL CSI acquisition. Furthermore, the Physical Broadcast Channel (PBCH), instrumental for broadcasting system information and aiding UEs in network access, also contributes to DL CSI estimation by UEs, acting as additional virtual pilots. This dense placement of virtual pilots (SSS, PSS, and PBCH) aids in detecting multipath effects with large delays, which CSI-RS might miss, despite the mismatch in bandwidth coverage with the bandwidth part (BWP) designated for UEs.

In an ideal scenario, combining the channels from sparse uniform pilots (CSI-RS) with those from dense virtual pilots would enable us to harness the strengths of both pilot types, leading to more accurate CSI recovery. However, the effectiveness of our proposed architecture, SRCsiNet, hinges on maintaining a uniform sampling relationship between input and output to exploit the Inverse Discrete Fourier Transform (IDFT) shifting invariance property.

This section will introduce the integration of a compressive sensing-based deep learning model into SRCsiNet, to address the challenges posed by a non-uniform pilot setup while effectively employing a bandpass filter. We will begin by outlining the compressive sensing-based CSI upsampling method, followed by an introduction to a novel framework, SRISTA-Net.

\subsection{Compressive sensing based CSI upsampling}
\begin{figure}
    \centering
    \resizebox{3in}{!}{
    \includegraphics{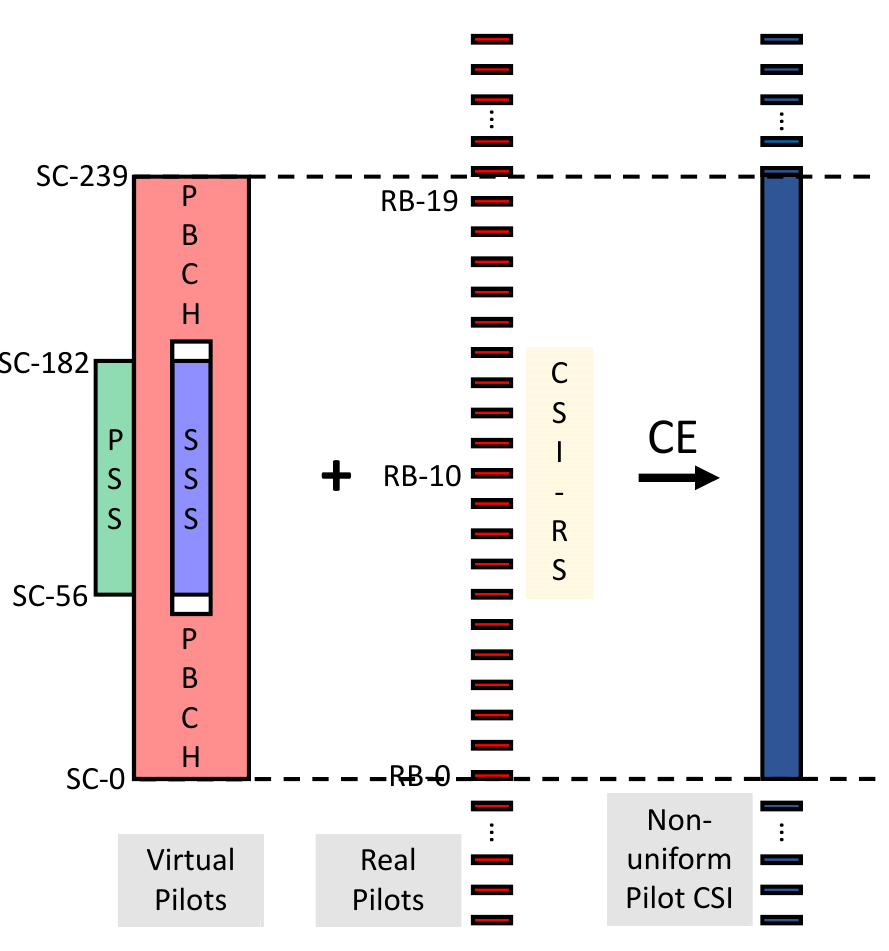}}
    \caption{Illustration of virtual pilots (i.e., PBCH, SSS and PSS) and non-uniform pilot DL CSI. With the sparse uniform pilots (CSI-RS) and the dense virtual pilots, we can have an effective non-uniform DL CSI.}
    \label{fig: virtual_pilots}
\end{figure}

As illustrated in Fig. \ref{fig: virtual_pilots}, considering the extra subcarrier-level DL CSIs, we can express the non-uniform pilot DL CSI, termed as LR DL CSI for simplicity, as
\begin{equation}
    \mathbf{H}_\text{LR}[i,j] = \left\{ 
    \begin{aligned}
    &\mathbf{H}[i,j], \forall j \in \Psi_P,\\
    &0, \forall j \notin \Psi_P,
    \end{aligned}
    \right.
\end{equation}
where $\Psi_P = \Psi_\text{RS} \cup \Psi_\text{ex}$ is the union of $\Psi_\text{RS}$ and $\Psi_\text{ex} = \{I, I+1,...,I+P-1\}$ with $I$ being the smallest subcarrier index in SSS, PSS or PBCH. $\Psi_\text{ex}$ is the index set of the consecutive pilots with size of $P$. We can reformulate the LR DL CSI based on the full AD DL CSI as 
%\begin{equation}
%    \begin{aligned}
%    \mathbf{F}_\text{BA}\mathbf{H}_\text{LR} =     \mathbf{F}_\text{BA}\mathbf{H}\mathbf{I}[:,\Phi_\text{P}] = \mathbf{F}_\text{BA}\mathbf{H}\mathbf{F}_\text{FD}\mathbf{F}_\text{FD}^{H}\mathbf{I}[:,\Phi_\text{P}] \\
    %=\mathbf{F}_\text{BA}\mathbf{H}_\text{AD}\mathbf{F}_\text{FD}^{H}\mathbf{I}[:,\Phi_\text{P}] = \mathbf{H}_\text{BD}\widetilde{\mathbf{F}}_\text{DF}, i = 1,..., N_a, 
%    \end{aligned}\label{eq: poi}
%\end{equation}

\begin{equation}
    \begin{aligned}
    \mathbf{H}_\text{LR} & = \mathbf{H}\mathbf{I}[:,\Phi_\text{P}] = \mathbf{H}\mathbf{F}_\text{FD}\mathbf{F}_\text{FD}^{H}\mathbf{I}[:,\Phi_\text{P}] \\
    &= \mathbf{H}_\text{AD}\mathbf{F}_\text{FD}^{H}\mathbf{I}[:,\Phi_\text{P}] = \mathbf{H}_\text{AD}\widetilde{\mathbf{F}}_\text{DF},
    \end{aligned}\label{eq: poi}
\end{equation}
where $\widetilde{\mathbf{F}}_\text{FD} = \mathbf{F}_\text{FD}[:,\Phi_\text{P}] \in \mathbb{C}^{N_f \times |\Phi_\text{P}|}$ is the trimmed DFT transformation matrix.

% represent H is function of H_LR, define the measurement matrix

Mathematically, the goal of compressive sensing reconstruction is to infer the original signal $\mathbf{x} \in \mathbb{C}^{N}$ from a low-dimensional measurement $\mathbf{y} = \mathbf{\Phi}\mathbf{x} \in \mathbb{C}^{M}$, where $M \ll N$. By transposing Eq.(\ref{eq: poi}), we have an exact projection of the problem of interest to a compressive sensing reconstruction problem (i.e., $\mathbf{y} = \mathbf{H}_\text{LR}[i,:]^T$, $\mathbf{\Phi} = \widetilde{\mathbf{F}}_\text{FD}^T$, $\mathbf{x} = \mathbf{H}_\text{AD}[i,:]^T$ where $i = 1,...,N_a$). This inversion is typically ill-posed problem. However, it can be solved by compressive sensing reconstruction since the sparsity of the original CSIs regularizes the possible outputs.

\subsection{ISTA-Net Framework}
Previous works have proposed a deep unfolding approach called ISTA-Net \cite{ISTANet}. The basic idea of ISTA-Net is to map the previous ISTA \cite{ISTA} approach updating steps to a deep learning network. This architecure consists of a fixed number of phases, each of the phase performs one iteration in classic ISTA algorithm.

\begin{figure*}
    \centering
    \resizebox{6in}{!}{
    \includegraphics{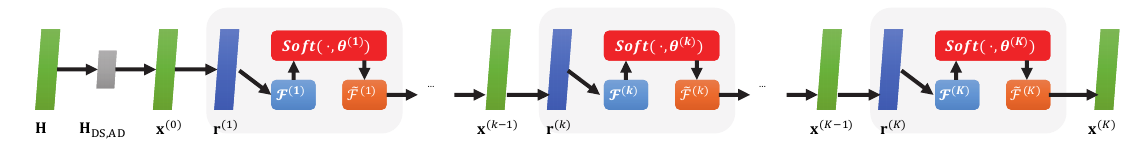}}
    \caption{Network architecture of ISTA-Net.}
    \label{fig: ISTANet}
\end{figure*}

Fig. \ref{fig: ISTANet} shows the deep learning network of the ISTA-Net. For each phase in ISTA-Net, it consists of two modules, namely the $\mathbf{r}^{(k)}$ \textbf{module} and the $\mathbf{x}^{(k)}$ \textbf{module}. The following items describe the operation in $k$-th phase as follows:
\begin{itemize}
    \item $\mathbf{r}^{(k)}$ \textbf{Module}: This aims to produce the intermediate result which is the same as the ISTA algorithm. This step is to optimize the channel fidelity $\norm{\widetilde{\mathbf{F}}_\text{FD}^T\mathbf{x}^{(k-1)}-\mathbf{H}_\text{LR}[i,:]^T}^2_2$. To maintain the ISTA architecture while increasing the channel similarity, a trainable step size $\rho^{(k)}$ to vary across different phases is adopted so that the output of this module with input $\mathbf{x}^{(k-1)}$ for $i$-th antenna can be represented as:
    
    \begin{equation}
        \begin{aligned}
        \mathbf{r}^{(k)} = \mathbf{x}^{(k-1)} - \rho^{(k)}\widetilde{\mathbf{F}}_\text{FD}(\widetilde{\mathbf{F}}_\text{FD}^T\mathbf{x}^{(k-1)} -\mathbf{H}_\text{LR}[i,:]^T).
        \end{aligned}
    \end{equation}
    
    \item $\mathbf{x}^{(k)}$ \textbf{Module}: It aims to compute $\mathbf{x}^{(k)}$ according to the intermediate result $\mathbf{r}^{(k)}$, which is given by
    \begin{equation}
        \mathbf{x}^{(k)} = \widetilde{\mathcal{F}}^{(k)}(soft(\mathcal{F}^{(k)}(\mathbf{r}^{k}),\theta^{(k)})), 
    \end{equation}
    where a pair of functions $\mathcal{F}^{(k)}$ and $\widetilde{\mathcal{F}}^{(k)}$ which are inverse of each other such that $\widetilde{\mathcal{F}}^{(k)}(\mathcal{F}^{(k)}(\cdot)) = \mathcal{I}(\cdot)$ with $\mathcal{I}(\cdot)$ being an identity function. Such a constraint on $\mathcal{F}^{(k)}$ and $\widetilde{\mathcal{F}}^{(k)}$ is called symmetry constraint.  
\end{itemize}

\subsection{SRISTA-Net Framework}
The ISTA-Net can deal with non-uniform sampling but cannot exploit side information. Thus, in this subsection, we propose a new framework which combines ISTA-Net and the proposed SRCsiNet for exploiting the advantages of the two networks, which is termed as \textit{SRISTA-Net}. 

\begin{figure*}
    \centering
    \resizebox{6in}{!}{
    \includegraphics{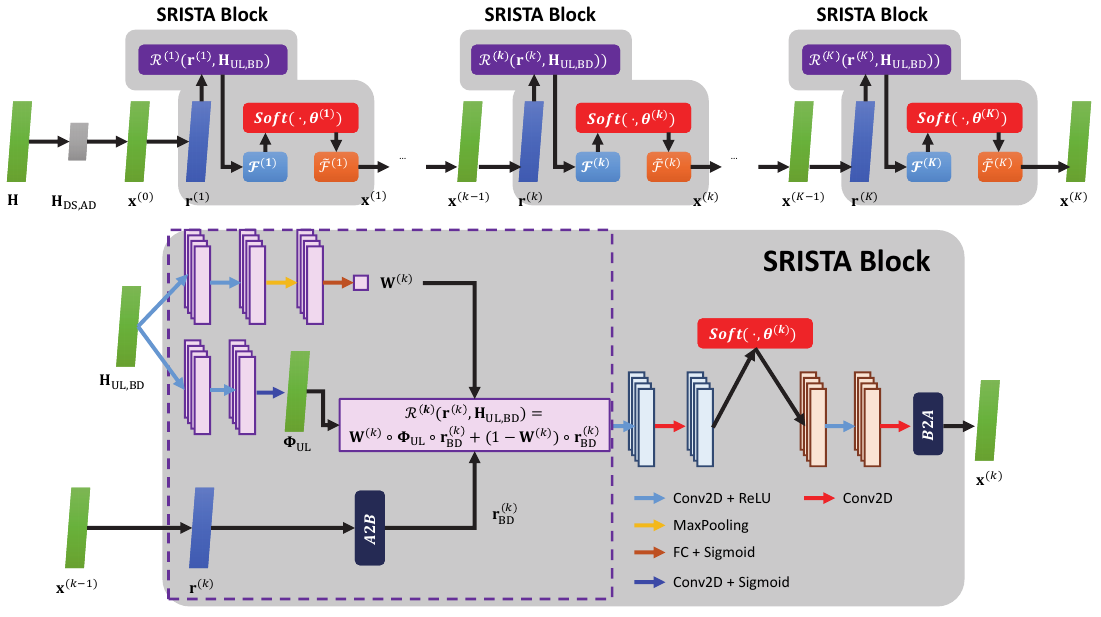}}
    \caption{Network architecture of SRISTA-Net. For the construction of $\mathbf{W}^{(k)}$, we employ a pair of 2D convolutional layers followed by max pooling operations. This approach is designed to refine the output, focusing it more acutely on specific segments of the side information. Subsequently, integrating a sigmoid layer as the terminal activation mechanism compels $\mathbf{W}^{(k)}$ to execute a binary fusion of the processed and unprocessed outcomes, specifically between $\mathbf{\Phi}_\text{UL}\circ\mathbf{r}^{(k)}_\text{BD}$ and $\mathbf{r}^{(k)}_\text{BD}$. As for the generation of $\mathbf{\Phi}_\text{UL}$, we apply BFD block in Eq. \ref{eq: BFD} mentioned in the previous section.}
    \label{fig: SRISTANet}
\end{figure*}

Fig. \ref{fig: SRISTANet} shows the deep learning network of the proposed network SRISTA-Net. We incorporate the SRCsiNet features into ISTA-Net by appending an additional block, Reciprocity Assisting (RA) Block, before the $\mathbf{r}^{(k)}$ module. This block aims to suppress the aliasing effects of the input $\mathbf{x}^{(k-1)}$ prior to solving the proximal mapping by applying the UL CSI assisted bandpass filter according to multipath reciprocity. We feed the magnitude of UL CSI $\mathbf{H}_\text{UL,BD}$ in the BD domain into two convolutional layers with ReLU and sigmoid functions, respectively, to obtain a bandpass filter $\mathbf{\Phi}_\text{UL}$.

Intuitively, for early phases, the model tends to heavily rely on UL CSI information and vice versa. Therefore, we design a weight matrix $\mathbf{W}^{(k)} \in \mathbb{C}^{N_a \times N_f}$ to adjust the dependency to the UL CSI at the $k$-th phase. We can rewrite the output of RA block as
\begin{equation}
    \mathcal{R}^{(k)}(\mathbf{r}^{(k)},\mathbf{H}_\text{UL,BD}) = \mathbf{W}^{(k)}\circ\mathbf{\Phi}_\text{UL}\circ\mathbf{r}^{(k)}_\text{BD} + (1-\mathbf{W}^{(k)})\circ\mathbf{r}^{(k)}_\text{BD},
\end{equation}
where $\mathbf{r}^{(k)}_\text{BD}$ is the $\mathbf{r}^{(k)}$ after transformation to BD domain. We then feed the output into the $\mathbf{x}^{(k)}$ module in ISTA-Net for minimizing the L1-norm constraints.

\subsection{Loss Function Design}
Given the training data pair $\{(\mathbf{H}_\text{DS}, \mathbf{H}_\text{UL,BD}, \mathbf{H})\}^{D}_{d=1}$, SRISTA-Net first transform $\mathbf{H}_\text{DS}$ into its AD version $\mathbf{H}_\text{DS,AD}$ as input and feed in the UL CSI information $\mathbf{H}_\text{UL,BD}$ in each phase to generate the output $\mathbf{x}^{(K)}_d$. Note that $\mathbf{H}_d$, $\mathbf{x}^{(k)}_d$ and $\mathbf{r}^{(k)}_d$ are all in the AF domain. To reduce the discrepancy between $\mathbf{H}_d$ and $\mathbf{x}^{(K)_d}$ while maintaining the symmetry constraint $\widetilde{\mathcal{F}}^{(k)}(\mathcal{F}^{(k)}(\cdot)) = \mathcal{I}(\cdot), \forall k = 1,...,K$, we design the following loss function:
\begin{equation}
    \mathcal{L}_\text{all}(\Theta) = \mathcal{L}_\text{discrepancy} + \gamma \mathcal{L}_\text{symmetry},
\end{equation}
\begin{equation}
    \mathcal{L}_\text{discrepancy} = \sum_{d=1}^{D}\norm{\mathbf{x}^{(K)}_d-\mathbf{H}_d}_2^2,
\end{equation}
\begin{equation}
    \mathcal{L}_\text{symmetry} = \sum_{d=1}^{D}\sum_{k=1}^{K}\norm{\widetilde{\mathcal{F}}^{(k)}(\mathcal{F}^{(k)}(\mathbf{q}^{(k)}_d))-\mathbf{q}^{(k)}_d}_2^2,
\end{equation}
where $\mathbf{q}^{(k)}_d = \mathcal{R}^{(k)}(\mathbf{r}^{(k)},\mathbf{H}_\text{UL,BD})$ is the output of the RA block at the $k$-th phase. $D$, $K$ and $\gamma$ are the total number of training data size, the total number of SRISTA-Net phases, and the regularization parameter, respectively. In this paper, we follow the original manuscript of ISTA-Net for the value of $\gamma = 0.01$.

\subsection{Initialization}
Like traditional iterative compressive sensing reconstruction, the proposed approach requires an initialization denoted by $\mathbf{x}^{(0)}$ as illustrated in Fig. \ref{fig: SRISTANet}. From Eq.(\ref{eq: poi}), we know $\mathbf{H}_\text{LR}[i,:]^T = \widetilde{\mathbf{F}}_\text{FD}^T\mathbf{H}_\text{AD}[i,:]^T, \forall i = 1,...,N_a$. We take the LS solution to this problem for initialization such that
\begin{equation}
    \mathbf{x}^{(0)} = \widetilde{\mathbf{F}}_\text{FD}^*(\widetilde{\mathbf{F}}^T_\text{FD}\widetilde{\mathbf{F}}_\text{FD}^*)^{-1}\mathbf{H}_\text{LR}^T
\end{equation}
\section{Experimental Evaluations}
\subsection{Experiment Setup}
Tests were focused on outdoor channels using widely used channel model software, QuaDriGa. The simulator considers a gNB with an $8 \times 4$ UPA and $32$-element ULA serving single-antenna UEs, respectively, with half-wavelength uniform spacing. 2000 UEs uniformly distribute in the cell coverage which is rectangular region with size of $250 (\text{m})\times$ 300 (\text{m}). The scenario features given in 3GPP TR 38.901 UMa were followed, using $N_f = 667$ subcarriers with $15K$-Hz spacing and $M_f = 55$ pilots with a downsampling ratio of $D_\text{RS} = 12$ as a common setting if not specified and assuming precise CSI estimates at the UEs. The NMSE metric was used to assess performance.

For DL-based models, we conducted training with a batch size of 32 for 1500 epochs, starting with a learning rate of 0.001 and setting an early stop criterion that validation loss does not improve for $100$ epochs. We generated the outdoor datasets using QuaDRiGa channel simulators. We consider $16$ TTIs for each out of 2000 UEs. In total, the dataset consists of 32,000 channels. We used one-tenth of the channels for testing and validation, respectively. The remaining four-fifths channels are  for training. 

For the ease to evaluate the degree of aliasing, it is common to use delay spread as a performance metric. A channel with larger delay spread tends to suffer aliasing effects more severely since it contains more high-delay multipaths. 
We cluster all the 3200 test CSI data into 3 clusters according to their RMS delay spread: low (smaller than 500 ns), medium (inbetween 500 ns and 1000 ns), high-delay spread (larger than 1000 ns). The low, medium and high delay spread clusters have 883, 1221 and 1095 test cases and are denoted as CL1, CL2 and CL3, respectively.

\subsection{UL Assisted Bandpass Filter Design for Anti-aliasing}
\begin{figure}
    \centering
    \resizebox{3.4in}{!}{\includegraphics{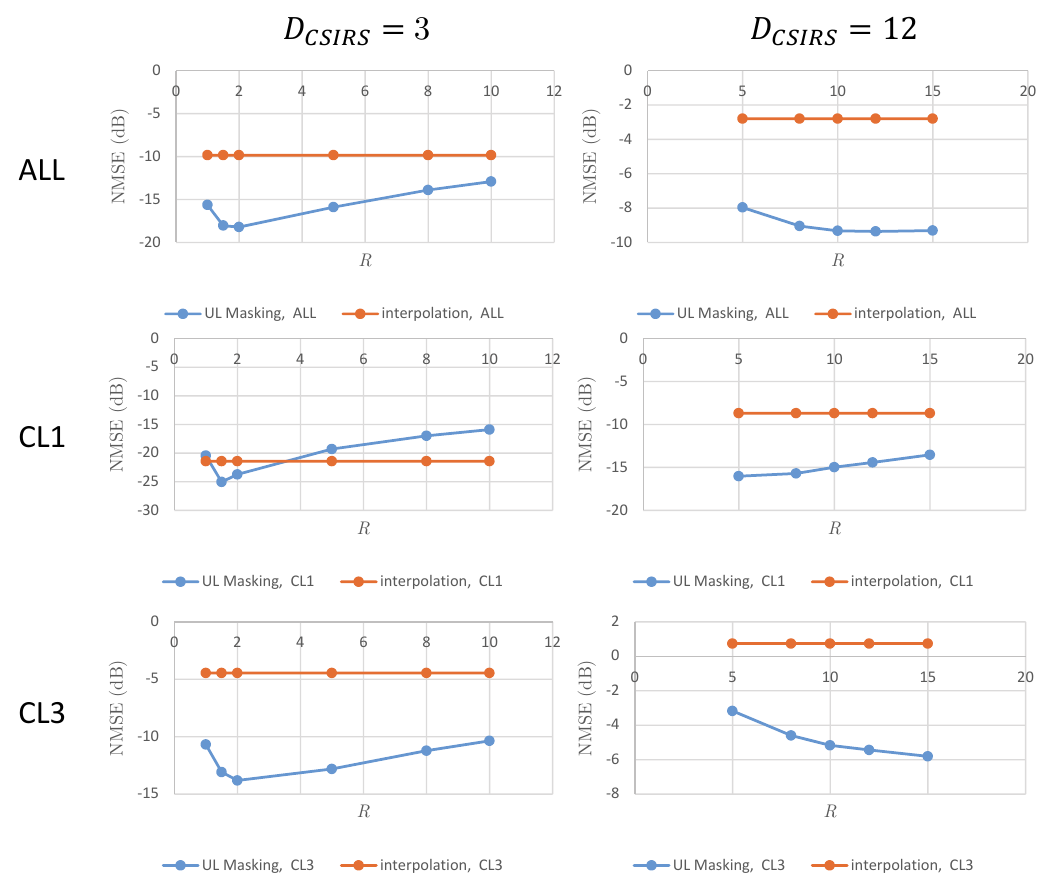}}
    \caption{NMSE performance of the proposed UL-assisted anti-aliasing and traditional linear interpolation for different CSI-RS placement densities ($D_\text{CSI-RS} = 3, 12$).}
    \label{fig: sim1-1}
\end{figure}

Fig. \ref{fig: sim1-1} displays the NMSE performance of the UL masking method at various $R$ levels compared to traditional interpolation across different CSI-RS placement densities. At a high CSI-RS density ($D_\text{RS} = 3$), the performance disparity between these approaches is minimal, notable mainly in the complete test dataset and CL1. However, a typical $D_\text{RS}$ value, being either 12 or 24, introduces a more significant aliasing effect. For $D_\text{RS} = 12$, the performance divergence becomes more pronounced, as the NMSE metrics show effective mitigation of aliasing effects, particularly in the high-delay-spread cluster, CL3.

\begin{figure*}
    \centering
    \resizebox{6.8in}{!}{\includegraphics{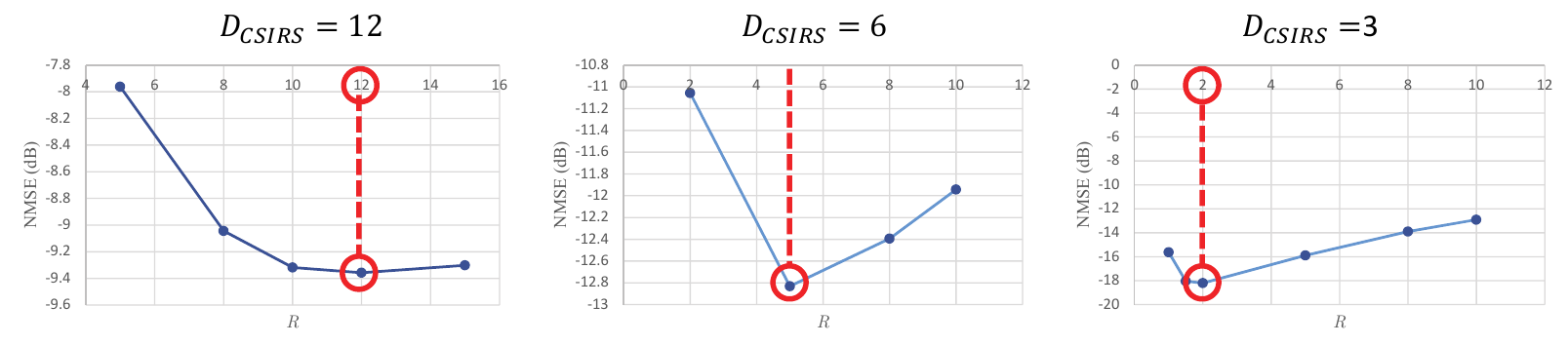}}
    \caption{NMSE performance of the proposed UL-assisted anti-aliasing for different CSI-RS placement densities ($D_\text{CSI-RS} = 12, 6, 3$). We can clearly know that the optimal selection of the threshold level $R$ varies with the aliasing effects. For the channels with strong aliasing effects, we require a larger $R$ to suppress aliasing copies.}
    \label{fig: sim1-2}
\end{figure*}

In Fig. \ref{fig: sim1-2}, the NMSE performance of the UL masking approach at varying $R$ levels for $D_\text{RS} = 3, 6, 12$ is depicted. This figure reveals the sensitivity of the proposed method to the choice of the UL masking parameter $R$. In cases of CSI with intense aliasing effects, a higher $R$ is necessary to effectively suppress the aliasing copies. Conversely, a large $R$ might be excessively aggressive for channels with a low delay spread, potentially compromising the integrity of the actual delay peaks.

\subsection{SRCsiNet}
\begin{figure}
    \centering
    \resizebox{3.4in}{!}{\includegraphics{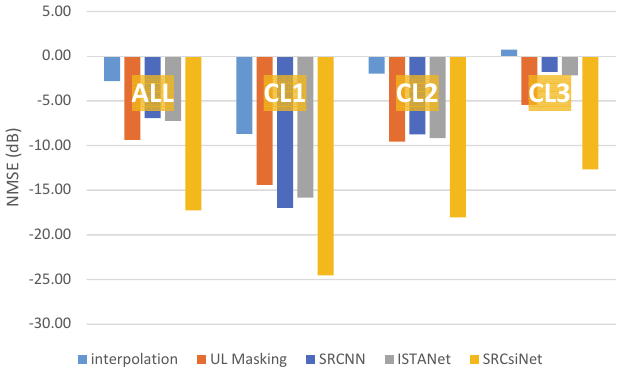}}
    \caption{NMSE performance of the learning based UL-assisted framework, SRCsiNet, and the alternatives in comparison for different clusters (i.e., all all samples and the samples with low, medium, large delay spread).}
    \label{fig: sim2-1}
\end{figure}

\begin{figure*}
    \centering
    \resizebox{6.8in}{!}{\includegraphics{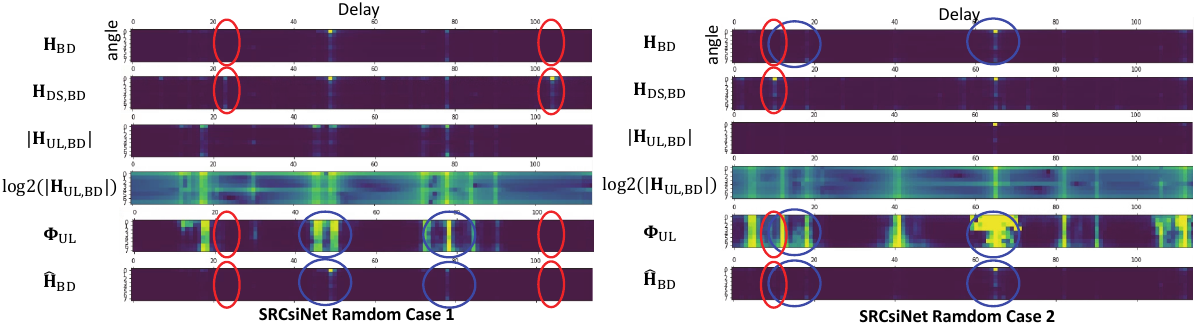}}
    \caption{Visual illustration of the results of the SRCsiNet. For the limited space, these examples only show the first 128 delay taps (we have 660 delay taps in the experiment). Since $D_{CSIRS} = 12$, $\mathbf{H}_\text{DS,BD}$ is periodic in every 55 delay taps. We can know from the examples that the bandpass filter works very well since it can capture very delicate details which are belong to true peaks (denoted by \textbf{blue} color circles) and suppress the aliasing peaks effectively (highlighted by \textbf{red} circles).}
    \label{fig: sim1-3}
\end{figure*}

In addition to the two upsampling approaches mentioned in the previous subsection, we compare them with the proposed learning-based SRCsiNet and SR network, SRCNN \cite{SRCNN} and a deep unfolding framework, ISTA-Net\cite{ISTANet}. Fig. \ref{fig: sim2-1} shows the NMSE performance of these alternatives for complete dataset and the three clusters. We can discover that ISTA-Net performs better than UL masking approach in CL1 due to the advantage of unfolding compressive sesning approach but performs poorly in CL3. That is because ISTA-Net does not introduce side information for dealing the aliasing effect. Clearly, by introducing UL CSI and providing flexibility in designing the bandpass filter, the overall performance can be improved by approximately 8 dB, which is significant. Fig. \ref{fig: sim1-3} shows the visualization of SRCsiNet. We can find that the bandpass filter design can effectively suppress the aliasing peaks and retain the delicate detail of the true peaks at the same time.

\subsection{End-to-end CSI Recovery}
In this subsection, we would like to demonstrate the importance of optimizing upsampling discrepancy for improving the overall performance. Table \ref{table: end2end} shows the NMSE performance from the end-to-end, feedback, and upsampling operation for SRISTA-Net, Interpolation and ISTA-Net. End-to-end NMSE performance would be bounded by either feedback loss or upsampling discrepancy. Yet, we can first discover that the end-to-end performance is generally bounded by upsampling loss in the considered UMa channels. This means that upsampling loss plays an critical role for improving the overall performance. Lastly, we can also find that the end-to-end NMSE performance improvement is about 6-10 dB as compared to other upsampling approaches without introducing UL CSI information.

\begin{table*}[]
\centering

\caption{The end-to-end NMSE performance of SRISTA-Net, Interpolation and ISTA-Net for different numbers of virtual pilots under compression ratio is 4.}
\begin{tabular}{|ccccc|c|ccccc|c|ccccc|}
\hline
\multicolumn{5}{|c|}{P   = 0}                                                                                                      &  & \multicolumn{5}{c|}{P   = 64}                                                                                                      &  & \multicolumn{5}{c|}{P   = 128}                                                                                                    \\ \hline
\multicolumn{5}{|c|}{DualNet-MP + SRISTA-Net}                                                                                    &  & \multicolumn{5}{c|}{DualNet-MP + SRISTA-Net}                                                                                     &  & \multicolumn{5}{c|}{DualNet-MP + SRISTA-Net}                                                                                    \\ \hline
\multicolumn{1}{|c|}{}              & \multicolumn{1}{c|}{ALL}   & \multicolumn{1}{c|}{CL1}   & \multicolumn{1}{c|}{CL2}   & CL3   &  & \multicolumn{1}{c|}{}              & \multicolumn{1}{c|}{ALL}   & \multicolumn{1}{c|}{CL1}   & \multicolumn{1}{c|}{CL2}   & CL3    &  & \multicolumn{1}{c|}{}              & \multicolumn{1}{c|}{ALL}   & \multicolumn{1}{c|}{CL1}   & \multicolumn{1}{c|}{CL2}   & CL3   \\ \hline
\multicolumn{1}{|c|}{$Loss$}    & \multicolumn{1}{c|}{-12.8} & \multicolumn{1}{c|}{-16.2} & \multicolumn{1}{c|}{-12.5} & -9.8  &  & \multicolumn{1}{c|}{$Loss$}    & \multicolumn{1}{c|}{-15.5} & \multicolumn{1}{c|}{-19.4} & \multicolumn{1}{c|}{-15.8} & -11.9  &  & \multicolumn{1}{c|}{$Loss$}    & \multicolumn{1}{c|}{-17.5} & \multicolumn{1}{c|}{-21.6} & \multicolumn{1}{c|}{-17.9} & -13.7 \\ \hline
\multicolumn{1}{|c|}{$Loss_\text{FB}$}       & \multicolumn{1}{c|}{-14.5} & \multicolumn{1}{c|}{-16.7} & \multicolumn{1}{c|}{-13.6} & -12.6 &  & \multicolumn{1}{c|}{$Loss_\text{FB}$}       & \multicolumn{1}{c|}{-19.6} & \multicolumn{1}{c|}{-23.2} & \multicolumn{1}{c|}{-19.7} & -16.3  &  & \multicolumn{1}{c|}{$Loss_\text{FB}$}       & \multicolumn{1}{c|}{-22.2} & \multicolumn{1}{c|}{-26.2} & \multicolumn{1}{c|}{-22.7} & -18.6 \\ \hline
\multicolumn{1}{|c|}{$Loss_\uparrow$}    & \multicolumn{1}{c|}{-17.2} & \multicolumn{1}{c|}{-24.5} & \multicolumn{1}{c|}{-18.0} & -12.6 &  & \multicolumn{1}{c|}{$Loss_\uparrow$}    & \multicolumn{1}{c|}{-17.6} & \multicolumn{1}{c|}{-22.3} & \multicolumn{1}{c|}{-18.7} & -13.4  &  & \multicolumn{1}{c|}{$Loss_\uparrow$}    & \multicolumn{1}{c|}{-19.4} & \multicolumn{1}{c|}{-24.0} & \multicolumn{1}{c|}{-20.1} & -15.3 \\ \hline
\multicolumn{5}{|c|}{DualNet-MP + Interpolation}                                                                               &  & \multicolumn{5}{c|}{DualNet-MP + Interpolation}                                                                                &  & \multicolumn{5}{c|}{DualNet-MP + Interpolation}                                                                               \\ \hline
\multicolumn{1}{|c|}{}              & \multicolumn{1}{c|}{ALL}   & \multicolumn{1}{c|}{CL1}   & \multicolumn{1}{c|}{CL2}   & CL3   &  & \multicolumn{1}{c|}{}              & \multicolumn{1}{c|}{ALL}   & \multicolumn{1}{c|}{CL1}   & \multicolumn{1}{c|}{CL2}   & CL3    &  & \multicolumn{1}{c|}{}              & \multicolumn{1}{c|}{ALL}   & \multicolumn{1}{c|}{CL1}   & \multicolumn{1}{c|}{CL2}   & CL3   \\ \hline
\multicolumn{1}{|c|}{$Loss$}    & \multicolumn{1}{c|}{-2.7}  & \multicolumn{1}{c|}{-8.3}  & \multicolumn{1}{c|}{-1.9}  & 0.7   &  & \multicolumn{1}{c|}{$Loss$}    & \multicolumn{1}{c|}{-3.2}  & \multicolumn{1}{c|}{-8.9}  & \multicolumn{1}{c|}{-2.3}  & 0.2    &  & \multicolumn{1}{c|}{$Loss$}    & \multicolumn{1}{c|}{-3.6}  & \multicolumn{1}{c|}{-9.4}  & \multicolumn{1}{c|}{-2.8}  & -0.1  \\ \hline
\multicolumn{1}{|c|}{$Loss_\text{FB}$}       & \multicolumn{1}{c|}{-14.5} & \multicolumn{1}{c|}{-16.7} & \multicolumn{1}{c|}{-13.6} & -12.6 &  & \multicolumn{1}{c|}{$Loss_\text{FB}$}       & \multicolumn{1}{c|}{-19.6} & \multicolumn{1}{c|}{-23.2} & \multicolumn{1}{c|}{-19.7} & -16.3  &  & \multicolumn{1}{c|}{$Loss_\text{FB}$}       & \multicolumn{1}{c|}{-22.2} & \multicolumn{1}{c|}{-26.2} & \multicolumn{1}{c|}{-22.7} & -18.6 \\ \hline
\multicolumn{1}{|c|}{$Loss_\uparrow$} & \multicolumn{1}{c|}{-2.7}  & \multicolumn{1}{c|}{-8.6}  & \multicolumn{1}{c|}{-1.9}  & 0.7   &  & \multicolumn{1}{c|}{$Loss_\uparrow$} & \multicolumn{1}{c|}{-3.2}  & \multicolumn{1}{c|}{-9.0}  & \multicolumn{1}{c|}{-2.3}  & 0.3    &  & \multicolumn{1}{c|}{$Loss_\uparrow$} & \multicolumn{1}{c|}{-3.6}  & \multicolumn{1}{c|}{-9.5}  & \multicolumn{1}{c|}{-2.8}  & -0.1  \\ \hline
\multicolumn{5}{|c|}{DualNet-MP + ISTA-Net}                                                                                    &  & \multicolumn{5}{c|}{DualNet-MP + ISTA-Net}                                                                                     &  & \multicolumn{5}{c|}{DualNet-MP + ISTA-Net}                                                                                    \\ \hline
\multicolumn{1}{|c|}{}              & \multicolumn{1}{c|}{ALL}   & \multicolumn{1}{c|}{CL1}   & \multicolumn{1}{c|}{CL2}   & CL3   &  & \multicolumn{1}{c|}{}              & \multicolumn{1}{c|}{ALL}   & \multicolumn{1}{c|}{CL1}   & \multicolumn{1}{c|}{CL2}   & CL3    &  & \multicolumn{1}{c|}{}              & \multicolumn{1}{c|}{ALL}   & \multicolumn{1}{c|}{CL1}   & \multicolumn{1}{c|}{CL2}   & CL3   \\ \hline
\multicolumn{1}{|c|}{$Loss$}    & \multicolumn{1}{c|}{-6.7}  & \multicolumn{1}{c|}{-13.5} & \multicolumn{1}{c|}{-8.1}  & -1.9  &  & \multicolumn{1}{c|}{$Loss$}    & \multicolumn{1}{c|}{-13.3} & \multicolumn{1}{c|}{-18.3} & \multicolumn{1}{c|}{-14.2} & -9.0   &  & \multicolumn{1}{c|}{$Loss$}    & \multicolumn{1}{c|}{-14.3} & \multicolumn{1}{c|}{-19.5} & \multicolumn{1}{c|}{-15.4} & -10.0 \\ \hline
\multicolumn{1}{|c|}{$Loss_\text{FB}$}       & \multicolumn{1}{c|}{-14.5} & \multicolumn{1}{c|}{-16.7} & \multicolumn{1}{c|}{-13.6} & -12.6 &  & \multicolumn{1}{c|}{$Loss_\text{FB}$}       & \multicolumn{1}{c|}{-19.6} & \multicolumn{1}{c|}{-23.2} & \multicolumn{1}{c|}{-19.7} & -16.36 &  & \multicolumn{1}{c|}{$Loss_\text{FB}$}       & \multicolumn{1}{c|}{-22.2} & \multicolumn{1}{c|}{-26.2} & \multicolumn{1}{c|}{-22.7} & -18.6 \\ \hline
\multicolumn{1}{|c|}{$Loss_\uparrow$}      & \multicolumn{1}{c|}{-7.2}  & \multicolumn{1}{c|}{-15.8} & \multicolumn{1}{c|}{-9.1}  & -2.1  &  & \multicolumn{1}{c|}{$Loss_\uparrow$}      & \multicolumn{1}{c|}{-14.5} & \multicolumn{1}{c|}{-20.5} & \multicolumn{1}{c|}{-15.9} & -9.9   &  & \multicolumn{1}{c|}{$Loss_\uparrow$}      & \multicolumn{1}{c|}{-15.3} & \multicolumn{1}{c|}{-20.8} & \multicolumn{1}{c|}{-16.5} & -10.8 \\ \hline
\end{tabular}
\label{table: end2end}
\end{table*}

\subsection{Solving Overfitting problem}
The SRISTA-Net architecture, necessitating 0.2 million parameters, faces a significant challenge due to its size relative to the training data, often leading to overfitting issues. This subsection highlights the effectiveness of Data Augmentation (DA) in our approach. Table \ref{table: SRISTA-Net DA} presents the NMSE performance for varying numbers of virtual pilots, comparing scenarios before and after implementing DA. A major hurdle in deploying learning-based models at gNB is the acquisition of real CSI data. In our experiments, the training of the deep learning model utilized less than 30,000 data points. We observed that overfitting becomes a significant issue when relying solely on the original training dataset. To counter this issue, we implemented circular shifting, as suggested by \cite{Zhenyu}, on the original training data in the angle domain, effectively doubling the training dataset size. This augmentation was found to markedly enhance NMSE performance, demonstrating the benefits of increased training data.

\begin{table}[]
\centering
\caption{NMSE performance of the SRISTA-Net with and without data augmentation (DA).}
\begin{tabular}{|c|l|l|l|l|l|}
\hline
\multicolumn{1}{|l|}{P} & Method          & ALL    & CL1    & CL2    & CL3    \\ \hline
\multirow{2}{*}{0}      & SRISTA-Net      & -14.62 & -21.34 & -15.41 & -10.12 \\ \cline{2-6} 
                        & SRISTA-Net + DA & -16.88 & -23.15 & -17.73 & -12.43 \\ \hline
\multirow{2}{*}{256}    & SRISTA-Net      & -17.20 & -22.48 & -18.34 & -12.83 \\ \cline{2-6} 
                        & SRISTA-Net + DA & -20.55 & -23.89 & -20.81 & -17.18 \\ \hline
\end{tabular}
\label{table: SRISTA-Net DA}
\end{table}

\subsection{Temporal Sensitivity of SRISTA-Net}
SRISTA-Net significantly surpasses other alternatives in NMSE performance. However, it is important to note that previous experiments were conducted under the assumption that both CSI-RS and virtual pilots are present within the same time slot\footnote{It's assumed here that the CSI remains constant within the same time slot}. Table \ref{table: time_sensitivity} details the NMSE performance of SRISTA-Net, accounting for varying time gaps between CSI-RS and virtual pilots, alongside different counts of virtual pilots. Given the 10 ms periodicity of PBCH, PSS, and SSS, the maximum theoretical time difference between CSI-RS and virtual pilots is limited to under 5 ms. Our findings reveal that SRISTA-Net's performance is highly susceptible to even minimal time differences, such as 5 ms. Interestingly, the NMSE performance in scenarios with a 5-ms gap is observed to be inferior compared to cases without any virtual pilots. In conclusion, when CSI-RS and virtual pilots coexist in the same time slot, leveraging the additional information is beneficial. Otherwise, it is preferable to upscale the DL CSI without incorporating data from virtual pilots.

% \footnote{Note that a typical periodicity of the virtual pilots (i.e., PBCH, PSS, and SSS) is $10 ms$. A reasonable time difference would be less than 10 ms.}
\begin{table}[]
\centering
\caption{NMSE perfornace of SRISTA-Net for different time differences between CSI-RS and virtual pilots.}
\begin{tabular}{|ccccc|}
\hline
\multicolumn{5}{|c|}{{\color[HTML]{000000} P = 64}}                                                                                                                                                                                                                                             \\ \hline
\multicolumn{1}{|c|}{{\color[HTML]{000000} \begin{tabular}[c]{@{}c@{}}Time\\ Difference\end{tabular}}} & \multicolumn{1}{c|}{{\color[HTML]{000000} ALL}}   & \multicolumn{1}{c|}{{\color[HTML]{000000} CL1}}   & \multicolumn{1}{c|}{{\color[HTML]{000000} CL2}}   & {\color[HTML]{000000} CL3}   \\ \hline
\multicolumn{1}{|c|}{{\color[HTML]{000000} 0ms}}                                                         & \multicolumn{1}{c|}{{\color[HTML]{000000} -17.6}} & \multicolumn{1}{c|}{{\color[HTML]{000000} -22.3}} & \multicolumn{1}{c|}{{\color[HTML]{000000} -18.7}} & {\color[HTML]{000000} -13.4} \\ \hline
\multicolumn{1}{|c|}{{\color[HTML]{000000} 5ms}}                                                       & \multicolumn{1}{c|}{{\color[HTML]{000000} -13.6}} & \multicolumn{1}{c|}{{\color[HTML]{000000} -15.0}} & \multicolumn{1}{c|}{{\color[HTML]{000000} -14.1}} & {\color[HTML]{000000} -11.2} \\ \hline
\multicolumn{1}{|c|}{{\color[HTML]{000000} 10ms}}                                                      & \multicolumn{1}{c|}{{\color[HTML]{000000} -9.2}}  & \multicolumn{1}{c|}{{\color[HTML]{000000} -10.1}} & \multicolumn{1}{c|}{{\color[HTML]{000000} -9.5}}  & {\color[HTML]{000000} -7.4}  \\ \hline
\multicolumn{1}{|c|}{{\color[HTML]{000000} \begin{tabular}[c]{@{}c@{}}One-shot\\ P=0\end{tabular}}}    & \multicolumn{1}{c|}{{\color[HTML]{000000} -17.2}} & \multicolumn{1}{c|}{{\color[HTML]{000000} -24.5}} & \multicolumn{1}{c|}{{\color[HTML]{000000} -18.0}} & {\color[HTML]{000000} -12.6} \\ \hline
\multicolumn{5}{|c|}{P = 128}                                                                                                                                                                                                                                                                     \\ \hline
\multicolumn{1}{|c|}{\begin{tabular}[c]{@{}c@{}}Time\\ Difference\end{tabular}}                        & \multicolumn{1}{c|}{ALL}                          & \multicolumn{1}{c|}{CL1}                          & \multicolumn{1}{c|}{CL2}                          & CL3                          \\ \hline
\multicolumn{1}{|c|}{0ms}                                                                                & \multicolumn{1}{c|}{-19.40}                       & \multicolumn{1}{c|}{-24.0}                        & \multicolumn{1}{c|}{-20.1}                        & -15.3                        \\ \hline
\multicolumn{1}{|c|}{5ms}                                                                              & \multicolumn{1}{c|}{-11.7}                        & \multicolumn{1}{c|}{-12.5}                        & \multicolumn{1}{c|}{-11.9}                        & -10.3                        \\ \hline
\multicolumn{1}{|c|}{10ms}                                                                             & \multicolumn{1}{c|}{-6.2}                         & \multicolumn{1}{c|}{-6.6}                         & \multicolumn{1}{c|}{-6.3}                         & -5.5                         \\ \hline
\multicolumn{1}{|c|}{\begin{tabular}[c]{@{}c@{}}One-shot\\ P=0\end{tabular}}                           & \multicolumn{1}{c|}{-17.2}                        & \multicolumn{1}{c|}{-24.5}                        & \multicolumn{1}{c|}{-18.0}                        & -12.6                        \\ \hline
\end{tabular}
\label{table: time_sensitivity}
\end{table}

\subsection{Complexity and Storage Requirements}

Table \ref{table: complexity and storage} outlines the complexity and storage requirements of all previously mentioned approaches. It is observed that while SRISTA-Net and ISTA-Net have similar model sizes and required similar complexities, SRISTA-Net significantly surpasses ISTA-Net in terms of performance. However, this comparison also highlights a drawback of deep unfolding methods. Due to the recursive application of convolutional operations on full-size data, these models exhibit higher complexity relative to others. Fortunately, the upsampling module in these models is implemented at the gNB. Considering the demands of future AI-enhanced cellular systems, a gNB equipped with multiple GPUs is envisioned, enabling real-time operation of such complex models. Nonetheless, there is an ongoing need to reduce the complexity of deep unfolding approaches, potentially through techniques like pruning \cite{pruning1, pruning2} or other methods of model size reduction.

\begin{table}[]
\centering
\caption{Storage (PARA: model parameters) and complexity (FLOPs) comparison.}
\begin{tabular}{|l|l|l|}
\hline
              & PARA & FLOPs \\ \hline
Interpolation & 0    & 109K  \\ \hline
UL Masking    & 0    & 206K  \\ \hline
SRCNN         & 63K  & 55.5M \\ \hline
ISTA-Net      & 196K & 2G    \\ \hline
SRCsiNet      & 7K   & 3.1M  \\ \hline
SRISTA-Net    & 215K & 2.01G \\ \hline
\end{tabular}
\label{table: complexity and storage}
\end{table}

\section{Conclusions}
The paper addresses a key challenge in massive MIMO FDD systems: the acquisition of DL CSI at the base station (gNB), which is crucial for optimal performance. It identifies a significant issue in current systems, where the undersampling of CSI due to low-density pilot placement leads to aliasing effects, impairing CSI recovery. To deal with this issue, the paper proposes a novel CSI upsampling framework for gNB, designed as a post-processing tool to fill the gaps caused by undersampling. This framework utilizes the principles of the DFT shifting theorem and multipath reciprocity, employing UL CSI to reduce aliasing effects. Additionally, the paper presents a learning-based approach that combines the proposed algorithm with the ISTA-Net architecture, aiming to improve non-uniform sampling recovery. The paper reports that both the rule-based and the deep learning methods demonstrate superior performance over traditional interpolation methods and current advanced techniques.

\bibliography{references.bib}
\bibliographystyle{IEEEtran}
\end{document}